\documentclass[conference]{IEEEtran}
\IEEEoverridecommandlockouts
\usepackage{cite}
\usepackage{bm}
\usepackage{amsmath,amssymb,amsfonts}
\usepackage{algorithmic}
\usepackage[linesnumbered,ruled,vlined]{algorithm2e}
\usepackage{graphicx}
\usepackage{amsfonts}
\usepackage{booktabs}
\usepackage{siunitx}
\usepackage{textcomp}
\usepackage{subcaption}
\usepackage{xcolor}
\usepackage{graphicx}
\usepackage{multirow}
\usepackage{lipsum} % 用于生成示例文本
\usepackage[ruled,vlined]{algorithm2e}
\def\BibTeX{{\rm B\kern-.05em{\sc i\kern-.025em b}\kern-.08em
    T\kern-.1667em\lower.7ex\hbox{E}\kern-.125emX}}
\title{Multi-Modal Neural Radio Radiance Field for Localized Statistical Channel Modelling\\
\thanks{The work is supported by the National Key R$\&$D Program of China (Grant No. 2023YFB2904804), the National Natural Science Foundation of China under Grant 62301334 and the Guangdong Major Project of Basic and Applied Basic Research (No. 2023B0303000001) {\em (Corresponding author: Ye Xue)}}
}

\author{
\IEEEauthorblockN{Yiheng Wang$^{1,2}$, Shutao Zhang$^{1,3}$, Ye Xue$^{1,2\#}$,  Tsung-Hui Chang$^{1,4}$}
\IEEEauthorblockA{$^{1}$ Shenzhen Research Institute of Big Data, The Chinese University of Hong Kong-Shenzhen, Guangdong, China\\
$^{2}$School of Data Science, The Chinese University of Hong Kong-Shenzhen, Guangdong, China\\
$^{3}$School of Science and Engineering, The Chinese University of Hong Kong-Shenzhen, Guangdong, China\\
$^{4}$School of Artificial Intelligence, The Chinese University of Hong Kong-Shenzhen, Guangdong, China\\
}

}

\begin{document}

\maketitle

\begin{abstract}
This paper presents MM-LSCM, a self-supervised multi-modal neural radio radiance field framework for localized statistical channel modeling (LSCM) for next-generation network optimization. Traditional LSCM methods rely solely on RSRP data, limiting their ability to model environmental structures that affect signal propagation. To address this, we propose a dual-branch neural architecture that integrates RSRP data and LiDAR point cloud information, enhancing spatial awareness and predictive accuracy. MM-LSCM leverages volume-rendering-based multi-modal synthesis to align radio propagation with environmental obstacles and employs a self-supervised training approach, eliminating the need for costly labeled data. Experimental results demonstrate that MM-LSCM significantly outperforms conventional methods in channel reconstruction accuracy and robustness to noise, making it a promising solution for real-world wireless network optimization.
\end{abstract}


\section{Introduction}

As next-generation wireless networks demand seamless connectivity, optimizing parameters such as antenna configurations, beamforming strategies, and transmit power has become increasingly complex and widespread \cite{10155734}. A critical challenge in network optimization is the accurate and cost-effective assessment of performance. While drive tests (DTs) \cite{hapsari2012minimization} remain a widely used method, they are costly, environmentally unfriendly, and limited to evaluating only existing network configurations. This underscores the pressing need for system-level simulations with precise radio propagation modeling \cite{zhang2023statistical}. At the core of such simulations lies the channel model, which captures localized physical structures to facilitate reliable network optimization. However, traditional statistical models, such as geometry-based stochastic models (GBSM) \cite{wang2018survey}, lack environmental specificity, whereas deterministic methods like ray tracing \cite{he2018design} demand high-fidelity maps and computationally expensive solutions based on Maxwell’s equations, limiting their practicality.

In contrast, physics-based and data-driven localized statistical channel modeling (LSCM) \cite{zhang2023physics,wang2024neural} has emerged as a promising approach. LSCM plays a crucial role in wireless network optimization by providing statistical insights into the wireless environment \cite{10155734}. It estimates the channel angular power spectrum (APS) from beam-wise reference signal received power (RSRP) data, enabling network performance prediction under varying network parameters, such as antenna array azimuth and tilt. However, LSCM methods rely solely on radio-related RSRP data from individual geographical grids. This approach is constrained by insufficient prior knowledge of environmental details and spatial correlations, limiting the ability to predict channels in unexplored areas.

 Deep learning methods \cite{orekondy2023winert,bakshi2019fast} have recently gained traction as a data-driven alternative to traditional channel modeling, circumventing explicit assumptions about signal propagation. Among these, NeRF-based approaches, such as $	\text{NeRF}^2$ \cite{zhao2023nerf2}, have demonstrated remarkable potential in modeling indoor, location-specific channel data. By leveraging neural radiance fields (NeRF) \cite{mildenhall2021nerf}, these methods can learn the underlying channel signal. However, their applicability remains limited to controlled indoor environments, as they rely on the availability of labeled channel data. Moreover, existing NeRF-based approaches for channel modelling do not incorporate environmental information, overlooking critical details that influence signal propagation.

Incorporating multi-modal sensory information presents a promising solution to these challenges. Recent advances in sensor manufacturing now enable the simultaneous collection of LiDAR 3D point clouds and RSRP measurements at minimal additional cost. These point clouds provide rich geometric representations of the environment, allowing models to explicitly incorporate spatial structures that govern signal propagation \cite{sun2023define}. However, a significant challenge remains: the potential misalignment between radio-related RSRP data and environmental LiDAR point cloud data, which can hinder model performance. 

To address these challenges, we propose a self-supervised modality alignment method for LSCM, named after multi-modal LSCM (MM-LSCM), leveraging both radio-related RSRP data and environment-related LiDAR point cloud data. Our contributions are threefold: 1) We propose a novel multi-modal neural radio radiance network specifically tailored for the LSCM problem. This dual-branch neural architecture can separately model the \textit{hitting probability } aided by prior volume density information from raw point cloud data and direction-dependent complex radio channel signals through a spherical harmonics function. 2) We introduce an implicit coherence weight, named after \textit{stopping probability}, during the volume-rendering-based multi-modal synthesis phase to align radio propagation characteristics and environmental obstacle characteristics. This coherence weight plays an important role in both channel-relative signal rendering and depth of the environmental obstacle prediction. 3) We propose a self-supervised training approach to enhance the model by leveraging prior depth information from point cloud data, eliminating the need for costly labeled data collection.

\section{System Model}\label{sec:sys}
\begin{figure}[htbp]
% \vspace{-3mm}
    \centering
    \includegraphics[width=0.5\textwidth]{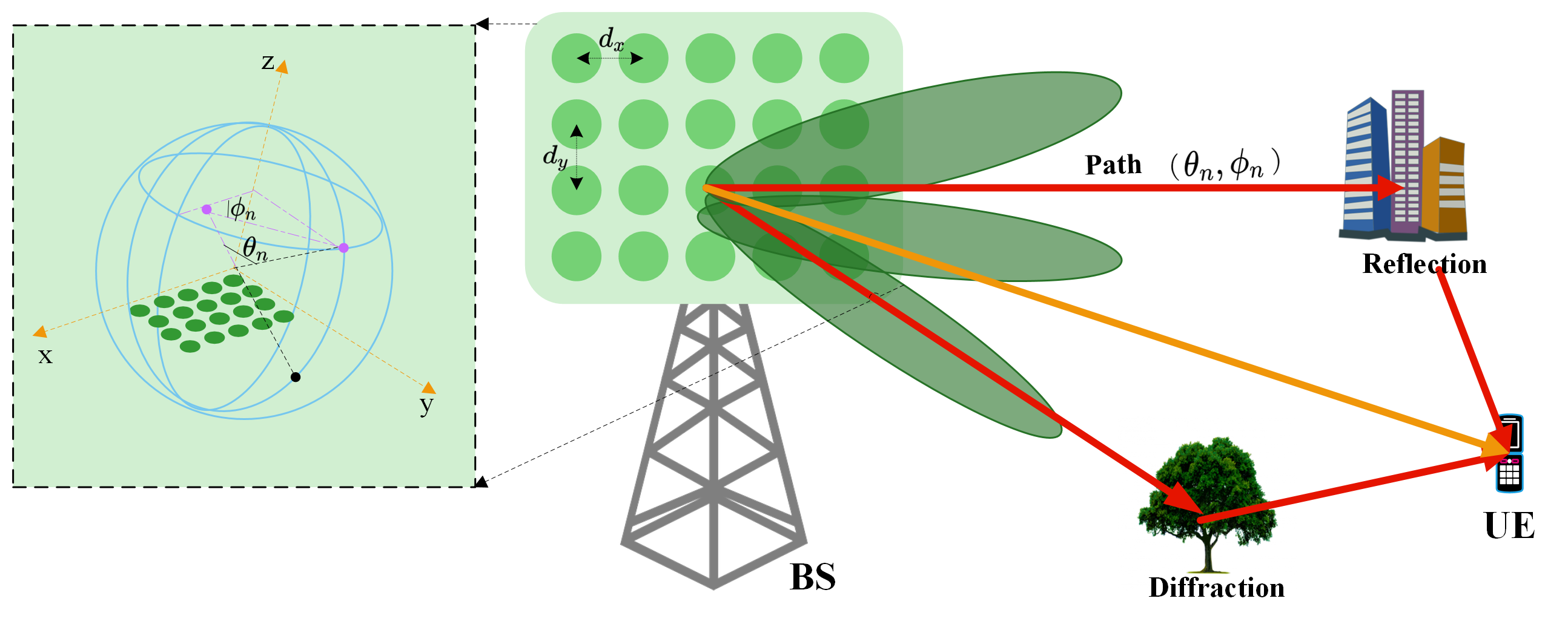}
    \caption{Angular Discretization of the Downlink Channel Model. The spherical coordinate is constructed relative to the UPA, where the $z$-axis points towards the normal direction of the UPA. The tilt AoD $\theta_n$ is defined as the angle between the $n$-th channel path and the $x$-$z$ plane. On the other hand, the azimuth AoD $\phi_n$ stands for the angle between the projection of the $n$-th channel path on the $x$-$z$ plane and the $z$-axis.} \label{fig:System Model}
    % \vspace{-5mm}
\end{figure}

We consider a downlink wireless system where one base station (BS) is equipped with an $N_T=N_x\times N_y$-antenna uniform planar array (UPA), as presented in Fig.~\ref{fig:System Model}. The BS transmits reference signals by using different $M$ beams from the codebook $\bm{W}=[\bm{w}_1,\ldots,\bm{w}_M]\in\mathbb{C}^{ N_T\times M}$ to sweep the angular space. The coverage region of the BS is divided into $L$ location grids.  The single-antenna user equipment (UE) at the $l$-th grid can receive the reference signal through the channel   $\bm{h}_{t,l}\in\mathbb{C}^{N_T}$ at the $t$-th sample, and then reports the RSRP to BS for LSCM, which focuses on the angle of departure (AoD) and channel gain
from BS to UEs \cite{ning2022multi}. Considering a widely used multi-path channel model with $\tilde{N}$ paths, and after discretizing the 3D space into $N\gg\tilde{N}$ space angles $\{(\theta_n,\phi_n)\mid  n\in[N]\}$
\cite{tse2005fundamentals} with $[N]\triangleq[1,2,\cdots,N]$, we have
\begin{equation}
    \bm{h}_{l,t}= \sum_{n=1}^N \bm{s}(\theta_{n},\phi_{n})\alpha_{n,l,t},\label{eq:angle}
\end{equation}
where  $\theta_{n}$, $\phi_{n}$, and $\alpha_{n,l,t}$ are the tilt AoD, azimuth AoD, and the corresponding complex channel gain for the $n$-th potential path, respectively. Note that if there does not exist a path in the discrete angle $\left( \theta_{n}, \phi_{n}\right)$, the corresponding channel gain $\alpha_{n,l,t}$ is zero. $\bm{s}(\theta_{n},\phi_{n})\in \mathbb{C}^{N_T}$ is the array steering vector, whose expression varies among different array geometries. Under our assumption, the array steering vector is defined as follows:
{
\begin{align}\label{steering}
	\bm{s}\left( \theta_{n}, \phi_{n} \right) =& \bm{s}_x\left( \theta_{n}, \phi_{n} \right) \otimes \bm{s}_y\left( \theta_{n} \right), \\
	\bm{s}_x\left( \theta_{n}, \phi_{n} \right) =& [1,  e^{-j\frac{ 2\pi}{\lambda}d_x\cos \theta_n\sin \varphi_n}, \cdots \nonumber \\
 &, e^{-j\frac{ 2\pi}{\lambda}\left( N_x - 1 \right)d_x\cos \theta_n\sin \phi_n}  ]^\textnormal{T}, \nonumber \\
	\bm{s}_y\left( \theta_{n} \right) =& \left[1,e^{-j\frac{2\pi}{\lambda}d_y\sin \theta_n }, \cdots,  e^{-j\frac{2\pi}{\lambda}\left( N_y - 1 \right)d_y\sin \theta_n } \right]^\textnormal{T} \nonumber.
\end{align}
} 

Specifically, the RSRP of the $m$-th beam reported from the $l$-th  grid at the $t$-th time  is given by 
\begin{equation}rsrp_{m,l,t}=P|\bm{h}^H_{l,t}\bm{w}_m|^2
\end{equation}

where  $P$ is the transmit power of the BS, which is typically known in advance. Thus, we normalize it to $1$ for simplicity. Without loss of generality, we assume that $\alpha_{n,l,t}$ are independent among different potential $n$-th paths, with phase following the uniform distribution over $[ -\pi,\pi]$. Let $\bm{rsrp}_{l,t}=[ rsrp_{1,l,t},\ldots,  rsrp_{M,l,t}]^\textnormal{T}$, and define the steering matrix $\bm{S}\in \mathbb{C}^{N_T\times N}$ by $\bm{S}=\left[\bm{s}(\theta_{1},\phi_{1}),\ldots, \bm{s}(\theta_{N},\phi_{N})\right]$. As derived  in \cite{zhang2023physics}, we can obtain the statistical relationship
\begin{equation}
    \mathbb{E}_{T}[\bm{rsrp}_{l,t}]=\big(|\bm{W}^\textnormal{H}\bm{S}|^2)\bm{x}_l,
\end{equation}
where the expectation $\mathbb{E}_T$ is taken over $T$ successive samples, $|\cdot|^2$  takes the element-wise absolute value and square. In addition, we can infer that $\bm{\alpha}_{l,t}=[ \alpha_{1,l,t},\ldots,  \alpha_{N,l,t}]^\textnormal{T}$ is the  APS  of channel, expressed as
% $\bm{x}_l=\mathbb{E}_{T}[\bm{\alpha}_{l,t}]$  
\begin{align}
    \bm{x}_l=\mathbb{E}_{T}[\bm{\alpha}_{l,t}]\in\mathbb R^{N}
\end{align}
remaining unchanged due to the static environment \cite{zhang2023physics}. Due to the limited number of scatterers and the stationarity of the environment,  $\bm{x}_l$ is usually sparse\cite{zhang2023physics}.  

After collecting the RSRP sample at the $l$-th grid over the total $T$ collected samples from DTs,  we can approximate the expected RSRP by computing the arithmetic mean of the RSRP collection. Therefore, we have 
\begin{equation}
    \bm{y}_l=\frac{1}{T}\sum_{t=1}^T\bm{rsrp}_{l,t}={\bm \Phi}\bm{x}_l+\bm{n}_l, \ \forall l\in[L],
\end{equation}
where we have ${\bm \Phi} \triangleq \big(|\bm{W}^H\bm{S}|^2)$ and $\bm{n}_l$  represents the residual noise, which accounts for approximation errors.

\section{Multi-Modal Localized Statistical Channel Modeling}
\begin{figure*}[t]
% \vspace{-3mm}
\centering
   \centering
    \includegraphics[width=0.85\textwidth]{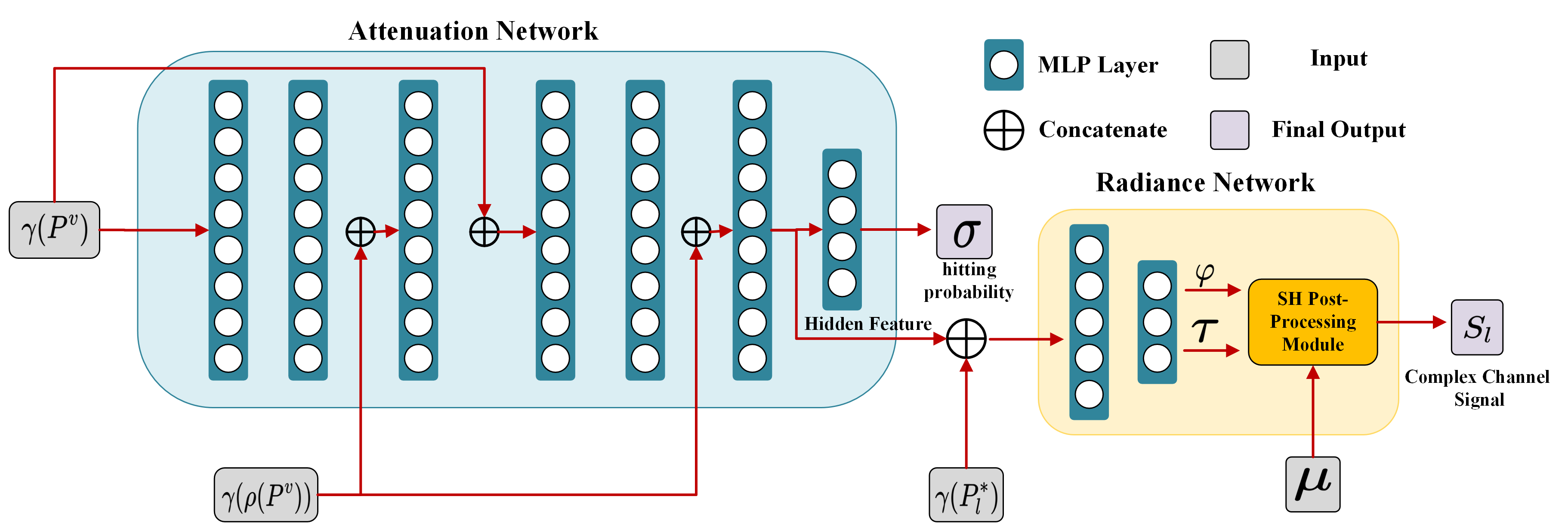}
    \caption{Illustration of Architecture of the Multi-Modal Neural Radio Radiance Network $\mathcal{F}_{\bm \Theta}$.}\label{fig:Network}
    % \vspace{-6mm}
\end{figure*}
In this section, we present MM-LSCM, a multi-modal neural radio channel representation framework designed to align radio-related RSRP data and environment-related LiDAR point cloud data for LSCM tasks.
\subsection{Problem Description}

Recovering all channel signal APSs across the BS's coverage grid $\{ l \in \mathbb{L}\}$ is challenging due to sparse and incomplete RSRP measurements, which are typically available only in limited explored regions $\{l \in \mathbb{L}_k \subseteq \mathbb{L}\}$. The remaining unexplored regions $\{l \in \mathbb{L}_o = \mathbb{L} \setminus \mathbb{L}_k\}$ lack direct RSRP measurements, necessitating predictive modeling for the full-channel APS $\{\bm{x}_l \mid l \in \mathbb{L}\}$. To address this, MM-LSCM utilizes sparse RSRP data $\{\bm{y}_l \mid l \in \mathbb{L}_k\}$, the 3D spatial coordinate of BS $P_{BS}$, and the measurement matrix $\bm{\Phi}$ to estimate APS values across all grid locations $\{\hat{\bm{x}}_l \mid l \in \mathbb{L}\}$ by leveraging the environmental features derived from raw 3D point clouds data $\bm P_{raw}=\{o_k=(x_k,y_k,z_k)\in \mathbb R^{3}|k\in[K]\}$ with $K$ numbers, collected from the coverage area of the BS to achieve its spatial alignment with radio-related data.

\subsection{Environmental Feature Design based on Spatial Voxelization}\label{sec:1}

 Firstly, we can create a bounded rectangular space, setting the BS position $P_{BS}$ as the origin, to cover all translated 3D point cloud data $\bm P_{raw}^{*}=\{o_k^{*}=o_k-P_{BS}|o_k\in \bm P_{raw}\}$ and translated grid locations $\{P^*_l=P_l-P_{BS}|l\in\mathbb L\}$ where $P_l$ is the original position of the $l$-grid. Then, to design the environmental features from 3D point cloud, we can discretize the scene of interest, w.r.t. the coverage region of BS, into a finite number of small, non-overlapping 3D cubic voxels, each characterized by a spatial coordinate ${P}^{v} \in \mathbb{R}^3$. This process partitions the original bounded rectangular space into a grid of dimensions $X \times Y \times Z$. After voxelization, we treat the location of obstacles as the environmental feature since it is important for radio propagation and can be retrieved from the 3D point cloud data. Specifically, we first calculated the volume density tensor $\bm\rho\in \mathbb{R}^{X \times Y \times Z}$ of a voxel with coordinate ${P}^{v}=(x_v,y_v,z_v)$ as 
\begin{align}
\bm\rho(x_v,y_v,z_v)\triangleq\sum_{k=1}^{K}\mathbb I\left(\begin{array}{c}
        \lfloor \frac{x^*_k - x_{\min}}{\Delta_x} \rfloor = x_v   \\
          \lfloor \frac{y^*_k - y_{\min}}{\Delta_y} \rfloor = y_v\\
          \lfloor \frac{z^*_k - z_{\min}}{\Delta_z} \rfloor = z_v
    \end{array}\right),\nonumber
\end{align}
where $(x^*_k,y^*_k,z^*_k)$ is the value of the $k$-th translated point cloud data $\bm P^*_{raw}=\{(x^*_k,y^*_k,z^*_k)\in \mathbb R^{3}\}^{K}_{k=1}$,
 $\mathbb{I}(\cdot)$ is an indicator function that evaluates to 1 when all three spatial coordinate conditions are met, and $\lfloor \cdot \rfloor$ denotes the flooring function. $\Delta_x, \Delta_y, \Delta_z$ are the voxel resolutions along each spatial axis and $x_{\min}, y_{\min}, z_{\min}$ represent the minimum coordinate values of $\bm P^*_{raw}$ along each axis. The density values then serve as ground truth indicators of whether an obstacle is present in voxels. Consequently, the location of major obstacles $P^{v*}$ can be obtained by identifying the nonzero density value $\bm\rho(P^{v*})$ closest to the BS since the major scatterers, formed by the obstacles, contributes to dominant signal paths, such as line-of-sight (LoS) and
strong non-line-of-sight (NLoS) reflections.
 
\subsection{Multi-Modal Neural Radio Radiance Network}
After spatial voxelization, we can consider each voxel as a basic element to describe the environment as well as a new radiance source retransmitting the RF signal according to the Huygens–Fresnel principle. This indicates that each voxel is encoded with its 3D spatial coordinate $P^v$,  the direction-dependent complex radio signal $S_{l}$ emitted from the voxel along the direction to the $l$-th grid, and its hitting probability $\sigma$, which is the probability of an electromagnetic wave hitting a particle at this voxel. Since $P^v$ can be easily obtained, we propose a multi-modal neural
radio radiance network to map the voxel coordinate $P^v$, its volume density feature $\bm{\rho}(P^v)$, and the translated target grid location $P^*_l$ to  $S_l$ and $\sigma$. In addition,  since the voxel may receive signals from different directions from the BS, which will influence its emitted radio signal $S_{l}$, we treat this direction, $\bm{\mu}$, as an input.
Specifically multi-modal neural
radio radiance network  is parameterized by learnable weights $\bm{\Theta}$ and expressed as
\begin{align} \mathcal{F}_{\bm \Theta}: \left(\gamma(P^v),\gamma\left(\bm\rho(P^v)\right),\gamma({P}^*_l),\bm \mu\right) \to (\sigma,S_l), \end{align}
where $\gamma(\cdot)$ is a position encoding function that enhances spatial resolution, defined as:
$ \gamma(P^v) = \big[\sin(\pi P^v), \cdots, \sin(\pi^V P^v), \cos(\pi^V P^v)\big] $
with encoding order $V$ \cite{tancik2020fourier}.

As shown in Fig.~\ref{fig:Network}, the multi-modal neural radio radiance network $\mathcal{F}_{\bm{\Theta}}$ adopts a dual-branch neural architecture consisting of an attenuation network and a radiance network. Unlike $\text{NeRF}^2$, the attenuation network is a residual MLP that incorporates the voxel’s point cloud density $\bm \rho(P^v)$ as an additional input to model hitting probability $\sigma$, which depends on the voxel’s spatial and environmental characteristics. Meanwhile, the radiance network is a fully connected MLP equipped with a post-processing module that applies the Spherical Harmonics (SH) function $\bm{sh}(\cdot)$ \cite{fridovich2022plenoxels} to model direction-dependent signals:
\begin{align}
    S_l=\bm{sh}(\bm \tau,\bm \mu)\odot e^{j\varphi}
\end{align}
where $\bm{\tau}$ and $\varphi$ are outputs of the fully connected MLP, representing the channel signal amplitude SH coefficients and the channel signal phase, respectively. $\odot$ denotes the Hadamard product. Furthermore, the multi-modal neural radio radiance network facilitates efficient volume-rendering-based multi-modal synthesis by enabling direct queries to derive the hitting probability and radiated channel signal at any voxel.

\subsection{Volume-Rendering-based Multi-Modal Synthesis}

After obtaining each voxel's corresponding radio and environment properties, we can synthesize the  APS and the depth of the environment obstacles by volume rendering.
% \BlankLine
\begin{figure}[htbp]
% \vspace{-5mm}
    \centering
    \includegraphics[width=0.45\textwidth]{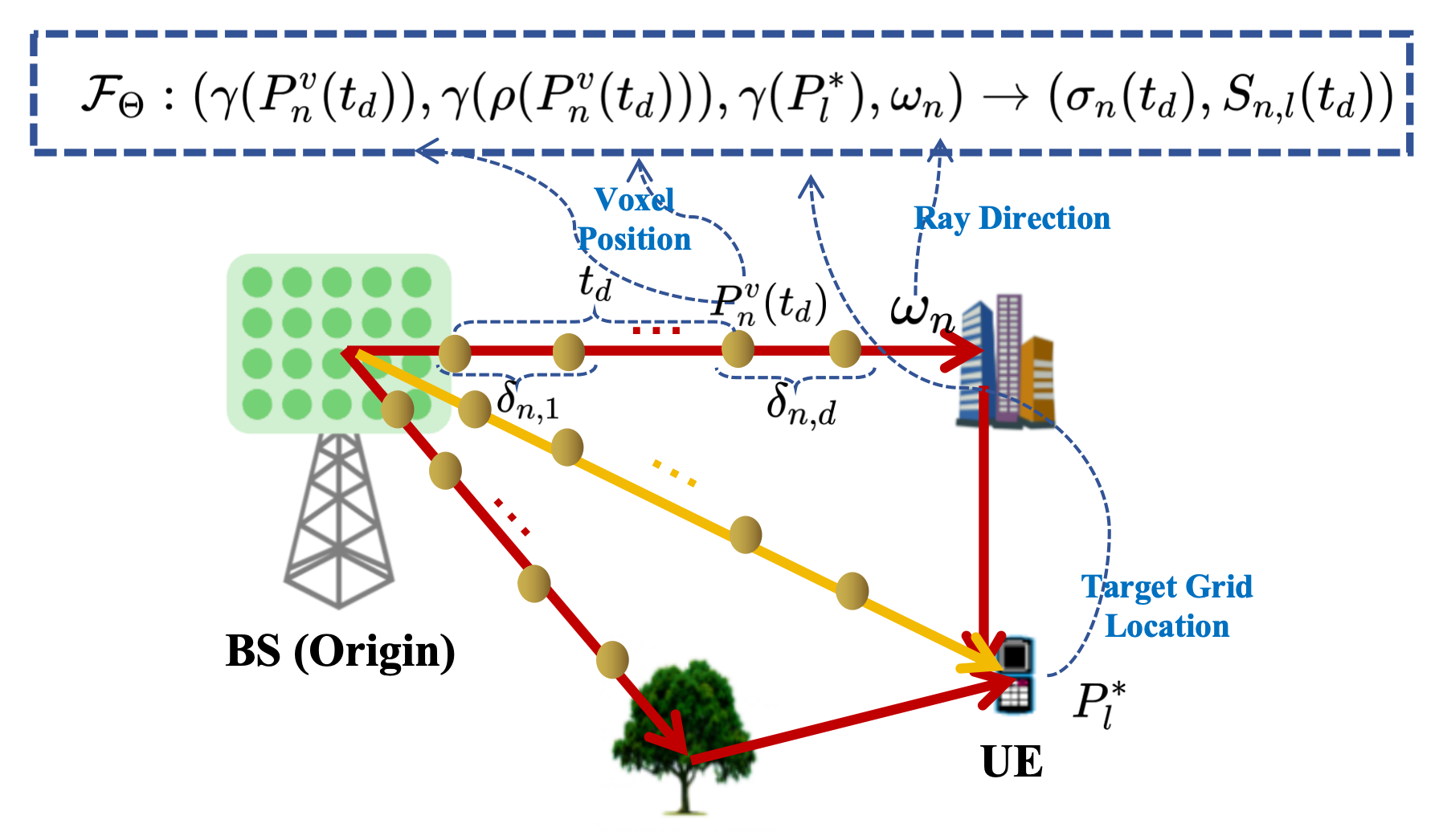}
    \caption{Illustration of Volume-Rendering-based Multi-Modal Synthesis.} \label{fig:ray}
% \vspace{-3mm}
\end{figure}

To enable efficient volume rendering to synthesize the  APS,  we discretize the space by rays starting from the BS, therefore, each voxel can be regarded as a point on a specific ray. Specifically, the position of voxel $P^v$ can be represented as \begin{align} P^v=P^v_{n}(t) \triangleq t \bm{\omega}_n,\label{eq:ray} \end{align} where $t$ represents the distance from the BS to the voxel. In addition, $\bm{\omega}_n$ is the unit vector represented the $n$-th direction from BS to voxel $P^v$, which can be represented by the AoD, $(\theta_n,\phi_n)$, of the $n$-th potential path  defined in Section \ref{sec:sys} as $\bm{\omega}_n= [\cos\phi_n\sin\theta_n, \sin\phi_n\sin\theta_n, \cos\theta_n]$. In this way, we can trace rays emitted by the BS  along the $n$-th ray to query the multi-modal neural radio radiance network $\mathcal{F}_{\bm{\Theta}}$ at voxel $P^v_{n}(t)$ with direction $\bm{\mu}=\bm{\omega}_n$ targeted to the APS of the $l$-th grid located at $P^*_l$. This enables access to the radiated channel signal $S_{l,n}(t)$ and the corresponding hitting probability $\sigma_{n}(t)$.

Then, we introduce the pass probability $\mathcal{T}_{n}(t_1,t_2)$ that quantifies the probability of the signal that propagates through some voxels without being absorbed or scattered along the $\bm \omega_n$ path from the voxel at $P^v_{n}(t_1)$ to the voxel at $P^v_{n}(t_2)$. Following the solution of  the differential equation in the volume rendering model in \cite{mildenhall2021nerf,tagliasacchi2022volume},  we have
\begin{align}
    \mathcal{T}_{n}(t_1,t_2)=\exp(-\int_{t_1}^{t_2} \sigma_{n}(t^\prime)dt^\prime),
\label{eq:intt}\end{align} which bridges the hitting probability $\sigma_{n}(t)$  and the pass probability. Using these two probabilities, we can describe the probability density for a wave stopping at a voxel along the $n$-th ray with distance $t$ as 
\begin{align} \nu_{n}(t)=\mathcal{T}_{n}(0,t)\cdot \sigma_n(t). \label{eq:nu} \end{align} This {\em stop probability} is crucial to synthesize both the APS and the depth of the environment obstacles since it serves as an implicit coherence weight, indicating the probability of a ray terminating at $P^v_{n}(t)$, where the radio signal are highly attenuated and the environmental obstacles are highly probable.

Specifically, to synthesize APS, we have the channel gain from the AoD of the $n$-th potential path to the $l$-th grid as
\begin{align} 
\hat{\bm{r}}_{l,n} &=\int_{0}^{t_{max}} \nu_{n}(t)\cdot S_{l,n}(t) dt,\label{eq:intaps}
\end{align}
where  $t_{max}$ is the distance of the farthest voxel. Then, the estimated APS can be obtained as
\begin{align} \hat{\bm{x}}_{l} = \left[ |\hat{\bm{r}}_{l,1}|^2, |\hat{\bm{r}}_{l,2}|^2, \dots, |\hat{\bm{r}}_{l,N}|^2 \right]^\textnormal{T}. \end{align}

To synthesize the depth of the environmental obstacles, we have 
\begin{align} \hat{z}_n&=\int_{0}^{t_{max}} \nu_{n}(t)\cdot t dt.\label{eq:expected}\end{align}

To avoid the high computation cost of calculating the continuous integrations (\ref{eq:intaps}) and (\ref{eq:expected}), we resort to approximating them by discrete summations. Specifically,  we approximate the continuous distance $t$ by a set of $D$ segments $\{\delta_{n,d}=t_{d+1}-t_d\}_{d=1}^{D}$ with $t_1=0, t_D=t_{max}$. In this way, the $n$-th element of APS can be derived from (\ref{eq:intt}) to (\ref{eq:intaps}), and the calculation of  definite integration as 
\begin{align}
    \hat{\bm{r}}_{l,n}\approx \sum_{d=1}^{D}  \mathcal{T}_{n,d}\cdot\mathcal{P}_{n,d}\cdot S_{l,n,d},\nonumber
\end{align}
where  we have \begin{align} \mathcal{T}_{n,d} &=\mathcal{T}_{n}(0,t_d)= \exp\left(-\sum_{d^{\prime}=1}^{d-1} \sigma_{n,d^{\prime}} \delta_{n,d^{\prime}}\right), \nonumber\\ \mathcal{P}_{n,d} &=\int_{t_d}^{t_{d+1}}\mathcal{T}_{n}(t_d,t)dt\cdot\sigma_{n,d}= 1 - \exp(-\sigma_{n,d} \delta_{n,d}),\nonumber\\
\sigma_{n,d}&=\sigma_{n}(t_d),\quad S_{l,n,d}=S_{l,n}(t_d), \end{align}  
\iffalse $(a)$ holds due to discrete voxel sampling, represented by $\mathbb{D} = \{P^v_{n,d} \mid P^v_{n,d}=P_{BS} + \sum_{d^\prime=0}^{d-1} \delta_{n,d^\prime} \bm{\omega}_n,  n \in [N],  d \in [D] \}$, with $\delta_{n,d}=t_{d+1}-t_d$ denoting the distance of $d$-th segment between consecutive sampled voxels with pairwise constant density $\sigma_{n,d}$ and radiated channel signal $S_{l,n,d}$, as illustrated in Fig.~\ref{fig:ray}. Here, 
where opacity $\mathcal{P}_{n,d}$ characterizes the degree to which the voxel $P^v_{n,d}$ absorbs or scatters the incident signal. Thus, the estimated APS at the target grid location $P_l$ can be obtained as:
\begin{align} \hat{\bm{x}}_{l} = \left[ |\hat{\bm{r}}_{l,1}|^2, |\hat{\bm{r}}_{l,2}|^2, \dots, |\hat{\bm{r}}_{l,N}|^2 \right]^\textnormal{T}, \end{align}
computed by aggregating the reconstructed signals $\hat{\bm{r}}_{l,n}$ along all rays $\mathbb{W}$. Moreover, inspired by \cite{rematas2022urban}, we observe that the voxel rendering integration weight $\nu_{n}(t)$ also serves as an implicit coherence weight, indicating the probability of a ray terminating at $P^v_{n}(t)$, where environmental obstacles are highly probable. Consequently, it can serve as a meaningful coherence weight, aligning radio propagation characteristics with environmental obstacles by guiding the expected depth $\hat{z}_n$ on $n$-th ray to match the depth from the LiDAR point cloud modality data: \fi 
Similarly, we can approximate the depth of the environmental obstacles as 
\begin{align} \hat{z}_n&\approx \sum_{d=1}^{D}  \mathcal{T}_{n,d}\cdot\left(-\exp\left(-\delta_{n,d}\sigma_{n,d}\right)\cdot t_{d+1}+t_d+\frac{\mathcal{P}_{n,d}}{\sigma_{n,d}}\right).\nonumber
\end{align}

\subsection{Optimizing MM-LSCM}

Given the challenges of obtaining labeled datasets for LSCM, we propose a novel self-supervised loss function to eliminate the need for ground-truth APS. Using the beam-wise RSRP collected over $T$ samples, the expected RSRP at $l$-th grid can be reconstructed from Eq.~(\ref{eq:angle}) as
\begin{align} \hat{\bm y}_l = \bm \Phi \hat{\bm x}_l, \quad \forall l \in \mathbb{L}. \end{align}
Thus, we learn $\bm{\Theta}$ by minimizing the following self-supervised loss over the limited measurement set $\{ l \in \mathbb{L}_k \subseteq \mathbb{L}\}$:
\begin{align} \ell_{radio} \triangleq \sum_{l\in\mathbb{L}_k} \|\bm{y}_l - \bm{\Phi} \hat{\bm{x}}_l\|_2^2 + \lambda_1 \|\hat{\bm{x}}_l\|_1, \label{eq:loss1} \end{align}
where $\|\cdot\|_1$ and $\|\cdot\|_2$ denote the $\ell_1$-Norm and $\ell_2$-Norm, respectively. The hyperparameter $\lambda_1$ enforces sparsity in $\hat{\bm{x}}_l$ through the $\ell_1$-Norm regularization. Notably, the training loss in Eq.~(\ref{eq:loss1}) relies solely on the available data points $\{\{\bm{y}_l\}_{l\in\mathbb{L}_k}, \bm{\Phi}\}$, eliminating the need for labeled APS data.

Moreover, since LiDAR point cloud modality data is available in our dataset, it can be leveraged to supervise model training.  To efficiently utilize the LiDAR point cloud, we represent it in a structured volume density tensor format $\bm \rho$ defined in Section \ref{sec:1}. Using Eq.~(\ref{eq:ray}), we trace all sampled voxels along each ray originating from the BS and query their corresponding density values, $\{\bm\rho(P^v_{n}(t_d))\mid n\in [N],  d\in [D]\}$. The volume density values serve as ground-truth indicators of whether an obstacle is present in each voxel. Consequently, the ground-truth depth of the environmental obstacles $z_n$ along the $n$-th ray is obtained by identifying the first nonzero density value $\bm\rho(P^{v}_{n}(t_{d^*}))$ along the ray $\bm \omega_n$:
\begin{align}
z_n=|P^{v}_{n}(t_{d^*})|=t_{d^*}.
\end{align}
To incorporate this information into model training, we supervise the expected depth $\hat{z}_n$, derived from the voxel rendering process in Eq.~(\ref{eq:expected}), to match the ground-truth depth obtained from the LiDAR point cloud:
\begin{align}
    \ell_{env}\triangleq\sum_{n=1}^{N}\| z_n-\hat{ z}_n\|_2^2. \label{eq:loss2}
\end{align}
By combining the radio modality loss and the environment-based supervision, we train the MM-LSCM with the objective:
\begin{align}
    \ell(\bm \Theta)\triangleq\ell_{radio}+\lambda_2\cdot\ell_{env}
\end{align}
Here, $\lambda_2$ is a hyperparameter that balances the contribution of the environment-related loss.
\section{Simulation Results and Discussions}
This section presents experiments showcasing the effectiveness and robustness of our proposed MM-LSCM when compared to prior and ablation baselines. All experiments were conducted on an NVIDIA A30 device using Python 3.8.19 with PyTorch 1.11.0.
\subsection{Setting of Training}

\begin{figure}[htbp]
\vspace{-8mm}
    \centering
    \includegraphics[width=0.30\textwidth]{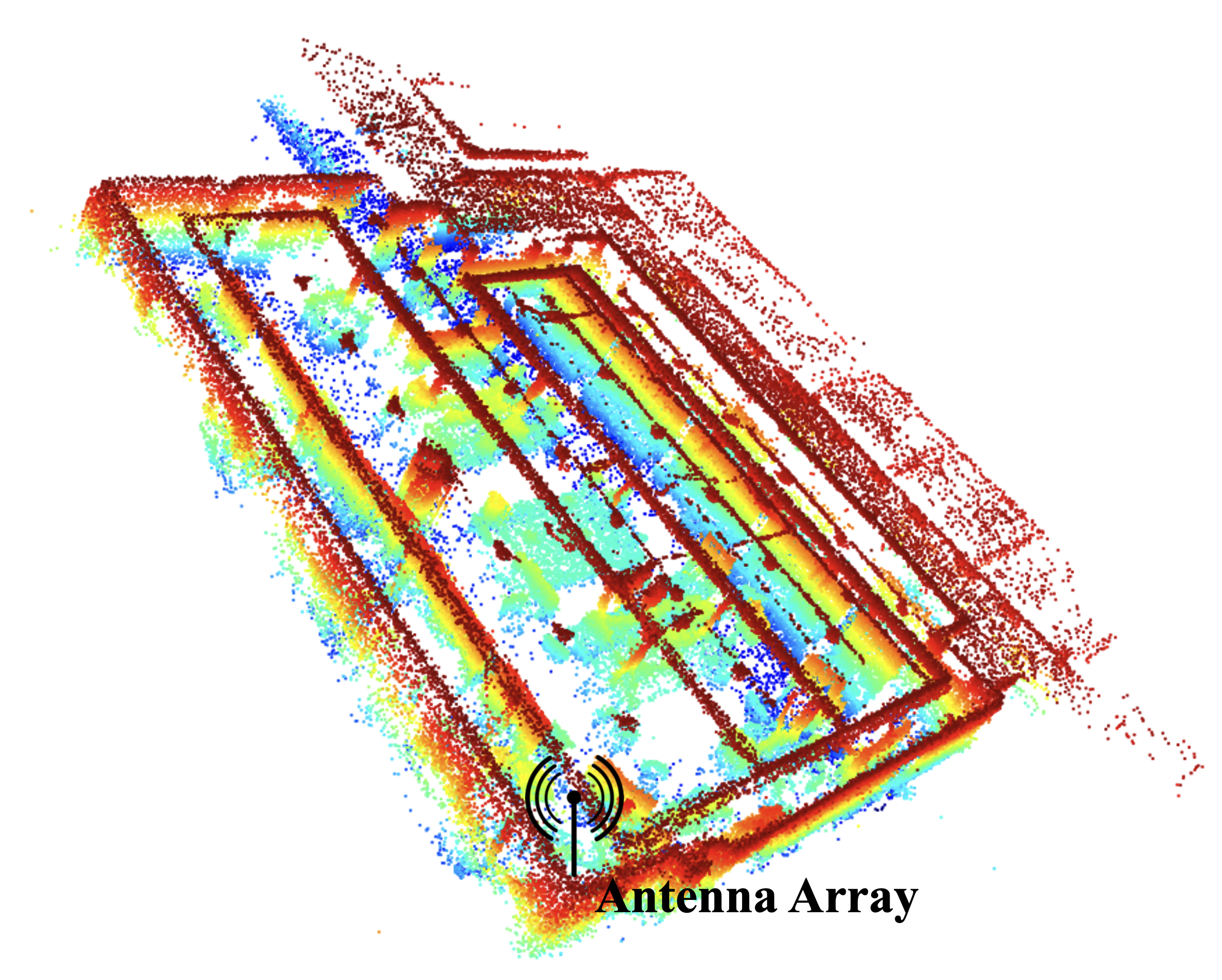}
    \caption{Illustration of Point Cloud of Laboratory Environment.} \label{fig:Indoor}
\end{figure}

We utilize the open-source real-world dataset provided by $\text{NeRF}^2$ \cite{zhao2023nerf2}, which consists of $K=192,416$ raw LiDAR point clouds, $\bm{P}_{raw}$, capturing a laboratory environment. This environment includes multiple tables, shelves, and several small rooms, with the corresponding 3D LiDAR point cloud representation shown in Fig.~\ref{fig:Indoor}. In this dataset, the receiver is equipped with a $4\times4$ antenna array operating at $915$ MHz, while the transmitter is an RFID tag continuously transmitting RN$16$ messages. The receiver is fixed at a corner of the laboratory, while the transmitter is systematically moved throughout the environment. 

To align with our task assumption, we consider the receiver as the BS and the transmitter as the UE. Consequently, the number of antennas in each dimension is $N_x = N_y = 4$. Each dataset sample consists of the UE's grid position $P_l$ and the corresponding spatial spectrum measured at the BS, located at $P_{BS}$. The spatial spectrum data can be downsampled into APS, denoted as $\bm{x}_l \in \mathbb{R}^{N}$, with $N=1620$\footnote{The original measured spatial spectrum is represented by $360\times90$ resolution from viewpoints sampled on the front hemisphere of the antenna array of the fixed receiver. Thus, we can downsample all of them into ones with $N=90\times18=1620$ as the angle space resolution on the front hemisphere of the antenna array of the BS.}. The dataset contains a total of $|\mathbb{L}|=6,123$ samples. For evaluation, we partition the dataset into an explored region, $\{P_l \mid l\in\mathbb{L}_k, |\mathbb{L}_k|=4,898\}$ (80\% of the data), and an unexplored region, $\{P_l \mid l\in\mathbb{L}_o, |\mathbb{L}_o|=1,225\}$ (20\% of the data).  To obtain the RSRP data, we commence by configuring the parameters of the downlink system in Fig.~\ref{fig:System Model} as follows: We set the number of antennas as $N_t =N_x\times N_y= 16$, the number of space angles as $N = 1620$, the number of beams as $M = 8$. We assume a UPA at the BS and employ a discrete Fourier transform (DFT) codebook matrix to construct the measurement matrix $\bm{\Phi} \in \mathbb{R}^{8\times1620}$. We generate the partially synthetic RSRP data:
\begin{align} \bm{y}_l = \bm{\Phi} \bm{x}_l, \quad \forall l \in \mathbb{L}. \label{eq:1}\end{align}
To simulate a multi-round scenario where the BS rotates its antennas, we apply once circular shift to the rows and columns of $\bm{\Phi}$ by $(\Delta_{\theta},\Delta_{\phi}) = (5^{\circ},4^{\circ})$, producing a rotated measurement matrix $\bm{\Phi}^r$. The rotated RSRP data samples are:
\begin{align} \bm{y}^r_l = \bm{\Phi}^r \bm{x}_l, \quad \forall l \in \mathbb{L}.\label{eq:2}\ \end{align}

Note that the training set only includes RSRPs data on the explored region $\{{\bm y}_l\mid l\in\mathbb{L}_k\}$, along with raw LiDAR point clouds $\bm P_{raw}$, the BS position $P_{BS}$ and grids' locations $\{{ P}_l\mid l\in\mathbb{L}_k\}$. To make use of all the partially synthetic data generated by Eq.~(\ref{eq:1}) and Eq.~(\ref{eq:2}), there are three sub-tasks to evaluate:
\begin{itemize}
    \item \textbf{Sub-Task 1} Predict the rotated RSRPs data $\{\hat{\bm y}_l^r=\bm\Phi^r\hat{\bm x}_l\mid l\in\mathbb{L}_k\}$ on the explored region $\{ l\in\mathbb{L}_k\}$.
    \item \textbf{Sub-Task 2} Predict the RSRPs data $\{\hat{\bm y}_l=\bm\Phi\hat{\bm x}_l\mid l\in\mathbb{L}_o\}$ on the unexplored region $\{ l\in\mathbb{L}_o\}$.
    \item \textbf{Sub-Task 3} Predict the rotated RSRPs data $\{\hat{\bm y}_l^r=\bm\Phi^r\hat{\bm x}_l\mid l\in\mathbb{L}_o\}$ on the unexplored region $\{ l\in\mathbb{L}_o\}$.
\end{itemize}
\subsection{Performance Comparison}
 In our experiments,  we consider WNOMP\cite{zhang2023physics} and SM-LSCM as our baselines. WNOMP is the SOTA method for recovering the APS from the corresponding RSRP in the LSCM task. It employs orthogonal matching pursuit-type algorithms to achieve APS reconstruction. SM-LSCM, adapted from $\text{NeRF}^2$\cite{zhao2023nerf2}, is a variant of the proposed MM-LSCM that does not incorporate point cloud data. Unlike MM-LSCM, SM-LSCM is trained solely by minimizing the radio modality loss (\ref{eq:loss1}). To evaluate performance, we use mean absolute error (MAE) and mean squared error (MSE) as key metrics:
\begin{align}
    \text {MAE} \triangleq\frac{1}{M}\|10\log\bm y_l-10\log\hat{\bm y}_l\|_1, 
    \text {MSE} \triangleq\frac{1}{N}\|\bm x_l-\hat{\bm x}_l\|_2^2.
\end{align} The results are shown in Table~\ref{tab:1}\footnote{Note that "{\bf --}" indicates that the WNOMP method is unable to handle the corresponding Sub-Task 2 and 3.}. They indicate that MM-LSCM consistently outperforms baselines across all sub-tasks. Besides, the performance gap between MM-LSCM and SM-LSCM highlights the significance of incorporating LiDAR-based point cloud data. These findings suggest that leveraging environmental information enables MM-LSCM to effectively align radio propagation characteristics with environmental obstacle characteristics, enhancing its predictive capabilities.

\begin{table}[htbp]
\centering
    \caption{Performance Comparison}
\resizebox{0.5\textwidth}{!}{%
\begin{tabular}{ccccccc}
\toprule
\hline
\multirow{2}{*}{\textbf{Method}} & \multicolumn{2}{c}{\textbf{Sub-Task 1}} & \multicolumn{2}{c}{\textbf{Sub-Task 2}} & \multicolumn{2}{c}{\textbf{Sub-Task 3}} \\ \cline{2-7} 
          & MAE (dB) & MSE      & MAE (dB)    & MSE         & MAE (dB)    & MSE         \\ \hline
MM-LSCM & 0.8666   & 0.0685   & 0.5628      & 0.0683      & 0.9017      & 0.0683      \\
WNOMP     & 1.8483   & 153.9319 & \textbf{--} & \textbf{--} & \textbf{--} & \textbf{--} \\
SM-LSCM   & 1.0923   & 12.1409  & 0.7013      & 12.1284     & 1.0996      & 12.1284     \\ \hline
\bottomrule
\end{tabular}%
}
\label{tab:1}
\end{table}

\begin{table}[htbp]
\vspace{-5mm}
\centering
    \caption{Robustness Comparison}
\resizebox{0.5\textwidth}{!}{%
\begin{tabular}{ccccccc}
\toprule
\hline
\multirow{2}{*}{\textbf{Method}} & \multicolumn{2}{c}{\textbf{Sub-Task 1}} & \multicolumn{2}{c}{\textbf{Sub-Task 2}} & \multicolumn{2}{c}{\textbf{Sub-Task 3}} \\ \cline{2-7} 
          & MAE (dB) & MSE      & MAE (dB)    & MSE         & MAE (dB)    & MSE         \\ \hline
MM-LSCM & 2.5861   & 0.0789   & 2.5842      & 0.0787      & 2.6329      & 0.0787      \\
WNOMP     & 4.0492   & 156.5648 & \textbf{--} & \textbf{--} & \textbf{--} & \textbf{--} \\
SM-LSCM   & 3.1984   & 14.2117   & 3.7867     & 14.2091     & 3.2287      & 14.2091     \\ \hline
\bottomrule
\end{tabular}%
}
\vspace{-6mm}
\label{tab:2}

\end{table}

\subsection{Robustness Comparison}
To further evaluate the robustness of our model, we introduce 3 dB noise to both the measurement matrix and the RSRP data. Specifically, we perturb two types of known measurement matrices and two types of RSRP data to simulate real-world conditions where the ground-truth measurement matrix is not perfectly known, and unavoidable noise is present in the collected RSRP data. The robustness comparison results, presented in Table~\ref{tab:2}, demonstrate that the trained MM-LSCM exhibits superior resilience to noise, enhancing its practicality for real-world applications.

\section{Conclusion}

In this paper, we present MM-LSCM, a self-supervised multi-modal neural radio radiance framework for LSCM. By integrating RSRP data with LiDAR-derived point cloud information, our model effectively captures environmental obstacles that influence radio signal propagation. Leveraging a dual-branch neural network and volume-rendering-based multi-modal synthesis, MM-LSCM achieves superior APS estimation compared to existing methods. Experimental results demonstrate improved predictive accuracy and robustness to noise, making MM-LSCM well-suited for real-world wireless network planning and optimization. Future work will explore the integration of additional modalities to further enhance performance for next-generation network optimization.

\bibliographystyle{IEEEtran}
\bibliography{refs.bib}
\end{document}